\documentclass[preprint2]{aastex}
\tightenlines 
\doublespace
\onecolumn

\newcommand{\etal}{{et~al.~}}

\newcommand{\coa}{\mbox{$^{13}$CO~}}
\newcommand{\cob}{\mbox{C$^{18}$O}}

\newcommand{\cc}{\mbox{${\rm cm}^{-3}$}}

\newcommand{\kms}{\mbox{${\rm km~s}^{-1}$}}

\newcommand{\vlsr}{\mbox{$\rm V_{\rm LSR}$~}}

\slugcomment{Submitted to The Astrophysical Journal}

\lefthead{Brunt \& Heyer}
\righthead{Interstellar Turbulence}

\begin{document}

\def\gtabouteq{\,\hbox{\raise 0.5 ex \hbox{$>$}\kern-.77em
                    \lower 0.5 ex \hbox{$\sim$}$\,$}}
\def\ltabouteq{\,\hbox{\raise 0.5 ex \hbox{$<$}\kern-.77em
                     \lower 0.5 ex \hbox{$\sim$}$\,$}}
\def\vlsr{V$_{LSR}$}
\def\kms{km s$^{-1}$}
\def\etal{{\it et al.}}

\title{Interstellar Turbulence:\\
     I. Retrieval of Velocity Field Statistics}

\author{Christopher Brunt$^{1}$}
\affil{Five College Radio Astronomy Observatory \& Department of Astronomy, Lederle Research Building,
University of Massachusetts, Amherst, MA 01003, USA}
\altaffiltext{1}{present address:
National Research Council, Herzberg Institute of Astrophysics, Dominion Radio Astrophysical Observatory, Penticton, BC, CANADA, and Department of Physics and Astronomy, University of Calgary, CANADA}
\and
\author{Mark H. Heyer}
\affil{Five College Radio Astronomy Observatory \&  Department of 
Astronomy, Lederle Research Building,
University of Massachusetts, Amherst, MA 01003, USA}

\begin{abstract}
We demonstrate the capability of Principal Component Analysis (PCA)
as applied by Heyer \& Schloerb (1997) to extract the statistics 
of turbulent interstellar velocity fields
as measured by the energy spectrum, $E(k) \sim k^{-\beta}$.  Turbulent 
velocity and density fields with known statistics are generated from 
fBm simulations.  These fields are translated to the observational 
domain, T(x,y,v), considering the excitation of molecular rotational
energy levels and radiative transfer.  Using PCA and 
the characterization of velocity and spatial scales from the
eigenvectors and eigenimages respectively, a relationship is identified 
which describes the magnitude of line profile differences, ${\delta}v$
 and the scale, $L$, 
over which these differences occur, ${\delta}v \sim L^\alpha$.
From a series of models 
with varying values of $\beta$,  we find,
$\alpha = 0.33\beta - 0.05$ for $1 < \beta < 3$.
This provides the basic calibration between the intrinsic 
velocity field statistics to observational measures and a diagnostic
for turbulent flows in the interstellar medium.  We also investigate the 
effects of noise, line opacity, and finite resolution on these results.
\end{abstract}
\keywords{hydrodynamics --- turbulence --- 
ISM: kinematics and dynamics --- ISM: clouds
line: profiles --- 
methods: statistical}

\clearpage
\section{Introduction}

The evolution of the cold, dense interstellar medium and the 
formation of stars are 
regulated by the kinetic energy content of turbulent gas flows.
Such chaotic 
motions 
 play a critical role in the equilibrium of the gas 
as these provide a countering pressure 
to self gravity and the pressure of the external medium.   
In the absence of external pressure fronts, 
the formation of stars 
occurs exclusively in dense, localized regions where the 
turbulent energy is sufficiently 
reduced to enable gravitational collapse. 
Therefore, an understanding of the dense interstellar medium and 
the formation of stars  requires 
an accurate description of turbulent gas flow including energy 
injection into and dissipation from the gas system and 
the distribution of turbulent kinetic energy over varying spatial 
scales.

Turbulent flows within the molecular 
interstellar medium are recognized as observed
supersonic line widths and morphological complexity (Scalo 1990; Falgarone
Phillips, \& Walker 1991).  While these observations denote the presence of turbulence 
in the interstellar medium, there is little quantitative information 
as to the nature of the turbulent flows.  
Ideally, one would like to
recover the 3-dimensional velocity field, $v_z(x,y,z)$,
for the line of sight component, z, 
from a set of spectroscopic observations.  In practice, the reconstruction
of the true velocity field 
from observations is not possible given the chaotic, non-deterministic
 nature of 
turbulent flows and the non-linear transformation
of the field onto the spectroscopic axis of the observer.  
Therefore, it is essential to construct statistical descriptions
of turbulence (Dickman 1985).
The primary statistic used
to distinguish turbulent models is the energy spectrum, E(k), which describes the 
degree of coherence of the velocity field over spatial scales.
In D dimensions,  the energy spectrum is the angular 
integral of the power spectrum,
$$ {E(k) = \int d\Omega P(k)  \propto <P(|k|)>_{\Omega} k^{D-1}} \eqno(1.1) $$
We refer to the spectral index $\beta$ in connection with
E(k), 
$$ {E(k) \sim |k|^{-\beta} }  \eqno(1.2) $$
The power spectrum then has the form (in D dimensions) :
$$ {P(k) \sim |k|^{-(\beta+(D-1))} } \eqno(1.3) $$
For a dissipationless
cascade of energy through an incompressible fluid,  
$\beta = 5/3$ and the power spectrum, P(k),
decreases in 3 dimensions as k$^{-11/3}$ (Kolmogorov 1941). 
A 2 or 1 dimension cut through the 3 dimensional structure will also
have an energy spectrum with the same index $\beta$ but with a
different slope of the power spectrum.

The mean square turbulent velocity, $v_l$,  at size $l$, 
is determined from the spectrum of velocity fluctuations
over scales smaller than $l$,
$$ v_l^2 \sim \int_{2\pi/l}^\infty E(k)dk 
        \propto \int_{2\pi/l}^\infty k^{-\beta}dk 
        \propto l^{\beta-1} \eqno(1.4)$$
$v_l^2$ describes the variance of the 
3 dimensional velocity field, v(x,y,z) within a volume $l^3$.  
The root mean square turbulent velocity, $<v_l^2>^{1/2}$, is then,
$$  <v_l^2>^{1/2} \propto l^{(\beta-1)/2} = l^\gamma \eqno(1.5) $$
where $\gamma=(\beta-1)/2$.
It is not equivalent  to 
an observed width of a spectral line which depends upon 
the velocity, density, temperature, and abundance which vary along the line
of sight. However, given
a velocity field described by an energy spectrum with $\beta > 0$, 
one might expect  some  
correlation of measured velocities, ${\delta}v$,
 with projected spatial scale, L, of the form,
$${\delta}v \sim L^\alpha \eqno(1.6)$$
where $\alpha$ refers to the measured spectral index in contrast to the 
intrinsic index $\gamma$ in equation (1.5).
Conventionally, ${\delta}v$ has been derived from measured line width
or the displacement of centroid velocities of spectral lines. 

Coarse limits to values of $\beta$ appropriate for the molecular interstellar
medium  can be inferred from the 
early models of observed CO line profiles.  
To account
for the broad velocity widths and the ability of the high opacity 
CO lines to detect the warm molecular gas located 
deep
within the Orion cloud, 
Goldreich \& Kwan 
(1974) modeled line profiles with a purely systematic velocity field 
replicating collapsing motions.  For such smooth, systematic velocity 
fields, $\beta \approx 3$.   However, it is unlikely that all molecular
clouds undergo global gravitational collapse since this predicts 
too large a star formation rate in the Galaxy (Zuckerman \& Evans 1974).
Other systematic velocity fields such as rotation are not observed
over the extended scales of molecular clouds (Arquilla \& Goldsmith 1984).
These results suggest $\beta$ must be smaller than 3.
A lower limit to $\beta$ is provided by the inability of purely random velocity 
fields ($\beta $ = -2 in 3D) to reproduce observed line profiles (Leung \& Lizst 1976; 
Dickman 1985).
Given these considerations, an approximate range to the spectral
index of the energy spectrum 
is
$ 3/2 < \beta < 5/2$.

More direct measures of $\beta$ rely on the identification of
a 
size-line width relationship and the assumption that variations 
along the spectroscopic
axis accurately reflect variations within the intrinsic velocity field.
Using a set of inhomogeneous data from the literature, 
Larson (1981) determined a relationship between the global velocity 
dispersion, $\sigma_v$, and size, L, for a sample of 
46 molecular regions in the 
solar neighborhood, $\sigma_v \sim L^{0.38}$.  Using a more homogeneous
sample of observations, Solomon etal (1987) identified a relationship
with an index of 0.5.  However, the applicability of these
observed relationships to the energy spectrum associated with interstellar 
turbulence (equation (1.5))  is weak
unless one considers the targeted clouds as 
part of singular, global fluid within the Galaxy (see Dickman 1985).  
Given the wide range of distances
and location of the clouds within different spiral arms, 
such an assumption is clearly
invalid.
Rather, these respective relationships derived from a
sample of unrelated clouds provide a statement of self gravitational 
equilibrium if one assumes that the mean column density of clouds 
is approximately constant.   That is, there is a sufficient amount 
of mass distributed within the cloud to bind the internal
turbulent motions.

Several investigations have decomposed 
spectroscopic data cubes of \coa~ or \cob~ emission from targeted molecular 
clouds into discrete units or ``clumps'' based on 
some preassigned definition (Carr 1987; 
Stutzki \& Gusten 1990; Williams, de Geus, \& Blitz 1994).  
For each clump, a global velocity dispersion and size are determined. 
However, the resultant size-line width relationships show a large scatter
and little significant correlation.  This may reflect that the localization
of emission within a spectroscopic data cube may not uniquely originate
from a discrete clump localized in  space.  Also, the process of identifying
clumps limits the resolution to the size of the smallest object.

Scalo (1984) and Kleiner \& Dickman (1985) proposed the use of 
structure and autocorrelation functions of centroid velocity images
derived from spectroscopic data cubes.  This assumes that 
the 2 dimensional projection
retains the statistical characteristics of the 3 dimensional velocity field.
These correlation methods have been applied 
by Miesch \& Bally (1994) to a sample of 12 molecular regions imaged 
in \coa~ emission.  The structure functions are described by  a power law
with a spectral index of 0.86
which corresponds to $\alpha$=0.43 as in equation (1.6).
However, these correlation
methods are limited.   
For many regions in the 
molecular ISM, observed line profiles exhibit non-gaussian shapes for which the 
centroid velocity is not an accurate approximation.  Also, the centroid 
velocity is not well defined near the edges of the clouds where the 
signal is weak.
Finally,
the relationship between 
the measured value of $\alpha$ and the intrinsic velocity field statistics,
$\gamma$ or $ \beta$, has 
not yet been established.

Recently, Heyer \& Schloerb (1997, hereafter HS97)  motivated the use
the multivariate technique 
of Principal Component Analysis (PCA) to 
decompose spectral line imaging observations of
molecular clouds.  
From the analysis, they identify 
a relationship which describes the magnitude of velocity 
differences of line profiles and the spatial scale at 
which these differences occur within the image.
For several molecular clouds, they find $\alpha \sim 0.4 - 0.5$.
However, it has not yet been demonstrated that this analysis,
or the other methods described above, are 
sensitive to varying field statistics ($\beta$) nor has 
the relationship between $\alpha$ and $\beta$ been established.
Given the complex transformation of the velocity field to the 
observed spectral axis, one can not assume that 
$\alpha = \gamma = (\beta-1)/2 $. 

In this paper, we quantitatively evaluate the ability of PCA to characterize 
intrinsic velocity fields from spectral line observations.
Guided by theoretical and observational results, 
3 dimensional  models of velocity and density fields 
with known statistical properties are generated. 
Observations of these fields are simulated, accounting for
non-LTE excitation and radiative transfer.  The simulated 
observations are analyzed with PCA under a variety of
physical conditions and observational limitations.  
We demonstrate that the method is indeed sensitive to 
velocity fields with varying energy spectra and 
derive the appropriate calibration between the 
observed quantity $\alpha$ and $\beta$.
In a companion paper (Brunt \& Heyer 2000),
we  use these results to derive the energy spectra for 
molecular regions in the outer Galaxy. 

\section{Construction of Model Turbulent Clouds}

\subsection{Velocity Field Generation}

To evaluate the sensitivity of the method outlined by HS97
to varying velocity 
field statistics, we construct 3 dimensional model 
velocity fields with 
prescribed energy spectra. 
Fields which exhibit varying degrees of spatial 
correlation can be generated from 
``colored
noise'' or fractional Brownian motion
(fBm) simulations (Voss 1988; Stutzki \etal~ 1998).
These simulations are not solutions to the fluid equations but rather,
are expedient numerical tools with well defined spatial statistics.
The resultant fields  are oversimplifications of 
real interstellar velocity fields in which one expects
shock structures, intermittency and  other
complex features.  

We use the simplest version of an
fBm {\it vector} velocity field, in which the components
are generated independently. For this type of field 
there is a roughly equal amount of 
power between the divergence free (solenoidal)
and the curl free (compressible) components. Our model
velocity fields are thus somewhat more extreme (in terms
of the energy fraction in the compressible modes) than
what may be expected for 'typical' turbulent fields which
tend to have their energy spectra dominated by the
divergence-free component. The effects, if any, of differing
energy fractions in each mode will be the subject of
a future paper. 

The model velocity field generation is a straightforward  process
which exploits the Fourier space
implementation of fBm in three dimensions. 
The three steps to generate the v(x,y,z) field are:
(1) distribute delta-correlated (Gaussian) noise on a cubic grid
of N $\times$ N $\times$ N pixels.
In this study, we use N=128;
(2) Fourier transform this field to the frequency domain, filter the Fourier
space field to the desired power law energy
spectrum of spectral index $\beta$, and transform back to
the real space domain;
(3) Normalize the field to the desired variance, $\sigma_v^2$. 
In this study, all velocity fields are normalized such that $\sigma_v^2=1$. 
Only one component of the
vector velocity field, the observable line-of-sight
component $v_{z}(x,y,z)$, is generated.
For our analysis, we consider purely isotropic fields. This means
that the longitudinal gradients $\delta v_{z}/\delta z$ are
statistically equivalent to the transverse
gradients ($\delta v_{z}/\delta x$ and
$\delta v_{z}/\delta y$).  

Ten realizations of v(x,y,z) fields are generated for each of several
values of $\beta$ (1.0, 1.5, 2.0 ,2.5, 3.0, 4.0).
The input noise for each realization has
a unique random number seed.
Representative one-dimensional cuts through fields with varying energy
spectra
generated by this method are displayed in Figure~1.
It is clear that as $\beta$ increases, the velocity field
becomes smoother and similar to a linear gradient. 

\subsection{Real-Space Velocity Field Statistics for Model Fields}

To ensure that the generated fields have the desired
real-space statistics (i.e. $\gamma$), 
we made direct (real-space) measurements
on the generated fields.  Of course, it would be possible 
to construct full-resolution autocorrelation functions or structure functions
but without the use of the Fourier Transform, these
would be lengthy calculations. We would like to avoid the
use of Fourier space in checking the fields, since these
were created within the Fourier domain. There is also the requirement
that the correct normalization for the measurements be
found.
To check the real-space statistics, we calculate a variant
of the standard (second order) structure function $S(\tau)$,
$$ S(\tau) = < (v(r) - v(r + \tau))^{2}>  \eqno(2.1)$$
where the angle brackets indicate averaging over all points
separated by lag $\tau$. We seek the desired dependence 
$$ S(\tau) \propto \tau^{2\gamma} \eqno(2.2) $$
$S(\tau)$ can be usefully
approximated
via coarse-grained versions of the original
field (Brunt 1999 and references therein).
Specifically, the field v is partitioned
into an ensemble of cubical boxes of L pixels on a side,
over which v is averaged. This procedure is carried
out for dyadic values of L (1, 2, 4, 8, 16, 32, 64), where
L=1 corresponds to the original field. For each
coarsened version of the field, the mean square difference
along the cardinal directions between {\it adjacent} coarse
pixels is calculated, providing an estimate of S($\tau$ = L).
The coarse structure function
$$ S_{C}(L) = < (v_{L}(\mbox{\boldmath $r$}) - v_{L}(\mbox{\boldmath $r + L$}))^{2} > \eqno(2.4) $$
where v$_{L}$ is a coarse-grained field. From these measurements
the exponent $\gamma$ is obtained via the proportionality 
$$ S_{C}(L) \propto L^{2\gamma} \eqno(2.5) $$

The {\it ensemble} coarse structure function, $<S_{C}(L)>$, is formed by
equal weight averaging of all S$_{C}$(L) measurements at
each L for each set of ten fields at each $\beta$.
These ensemble measurements are shown in Figure~2,
The deficit at L=64 
is partly caused by the periodicity of the
fields. 
A power law is fit to $<S_{C}(L)>$ for L $\leq$ 32 to determine an 
ensemble value, $\gamma_E$, for each $\beta$. 
Also, a mean value, $<\gamma>$ is derived from the average of
all $\gamma$ values obtained individually at each $\beta$.
These results 
are summarized in Table 1.  It is clear that the calculated values
of $\gamma$ are equivalent to the expected value $(\beta-1)/2$, 
derived
from the input value of $\beta$.
In Figure~3, we show the variation of $\gamma_E$ with the 
intrinsic value of $\beta$.
For velocity fields with $\beta \ge 3$, we find that 
$\gamma$ asymptotically approaches unity. 
This is due to the rapid decrease of the 
power spectrum such that only the
$|$k$|$ $= 1$ components are significant.  Therefore, the 
field is similar to a smooth linear gradient.  In a recent study,
Myers \& Gammie (1999) derive a similar conclusion.
The lower values of $\gamma_E$ and $<\gamma>$ relative to the 
expected values
for the larger $\beta$ values are likely caused by the
periodicity requirement, which suppresses the out-of-band
(large-scale) gradients, and also partly due to the gradients
which result from stochastic fluctuations of the power in
the first two wavenumbers. That is, an excess of power at the first
wavenumber cannot make the field more smooth (higher $\gamma$),
but a deficit of power at this wavenumber {\it can} make the
field less smooth (lower $\gamma$). 
A better approximation to  
``absolutely smooth'' fields requires the extension to
$\beta=4$.

\subsection{Density Field Generation}

We consider two forms of the density field.  The basic 
calibration is established with a uniform density field,
$n_\circ(x,y,z)$=constant.
However, to provide a more realistic observation, an
inhomogeneous  
density field is also considered. 
Since the spatial variation of density corresponds to varying excitation
of the trace molecule, these inhomogeneous models are used to 
to investigate the effects of
non-uniform sampling of the velocity field. 
Recent hydrodynamical models for an isothermal gas show
a lognormal density probability density function,
PDF 
(Padoan et al 1997; Passot \&
Vazquez-Semadeni 1998).
Additionally, the power
spectrum of the density distribution, in the simulations
of Padoan et al (1997), is consistent with a power-law form, 
of spectral slope (in the power spectrum) of $\sim 2.6$ and 
similar to values determined from observations (Stutzki \etal~ 1998).

To generate a field that has a lognormal PDF and
a power-law power spectrum, we implement a modified
version of the fBm method described in Section 2.1.
An fBm field 
with $\beta = 1$ is
exponentiated which produces a resultant field with the 
desired statistics
(Schertzer \&
Lovejoy 1987; Pecknold \etal~ 1993). The $\beta = 1$ fBm field 
serves as $\ln n $ + constant, where the normalizing
constant, given the dispersion in the field,
determines the mean density. The normalization of the
field to a mean density of $n_\circ = 1$ is obtained by setting
$<\ln n>  =  -\sigma_{\ln n}$, where $\sigma_{\ln n}$
is the global standard deviation of $\ln n$. Note that
$\sigma_{\ln n}$ is independent of $n_\circ$. 
Like the fBm velocity fields, the density field is
periodic across the cube boundaries. 
Therefore, the density field is inspected and, if necessary, re-centered
to ensure that no major structures lie across the
boundary. 
We emphasize that there is no spatial correlation between 
the density and velocity fields as one might expect for real clouds
in which density fluctuations result from the localized divergence 
of the velocity field.
 
\subsection{Physical Properties of the Models}

In order to calculate realistic opacities, we assign a 
physical scale to the simulation.
 For all realizations the linear size of the cube is 20pc.
A pixel or cell is 20pc/128=0.156 pc.
The mean density for the uniform density field is 100 cm$^{-3}$.
For the log normal filds, the  mean density varies from 
10$^{2}$ to 10$^{4}$ cm$^{-3}$. The gas
is assumed to be isothermal 
with a kinetic temperature,
T$_{k}$ equal to 20 K. 
An approximate measure of the global gravitational equilibrium 
of the models is
given by the ratio of kinetic ($\sim 3\sigma^{2}_{v_{z}}\rho_{0}r^{3}$)
to gravitational ($\sim G\rho_{0}^{2}r^{5}$) energies 
$$ 3\sigma^{2}_{v_{z}}/G\rho_{0}r^{2} \sim 
1.2 (100/<n_{H_{2}}>) \eqno(2.6) $$
where we have taken the radius, r= 10 pc. 
The models do not significantly deviate from
gravitational equilibrium although the equilibrium state of the model
clouds is not critical to these results.

\subsection{Observations of the Models} 

Given the density and velocity fields and the kinetic
temperature, the brightness temperature is determined from 
the excitation of the chosen gas 
probe and the radiative transfer of the emission toward the
observer's line of sight.  The excitation of the line, parameterized
by the excitation temperature, T$_{ex}$, and line center opacity,
$\tau_{\circ}$, at the 
position (x,y,z), are determined from a non LTE calculation which accounts
for local radiative trapping (Scoville \& Solomon 1974). This depends
on the local kinetic temperature, density, and N$_{H_{2}}$X/${\Delta}$v,
where N$_{H_{2}}$ is the molecular hydrogen column density, X is the 
abundance ratio to molecular hydrogen, and ${\Delta}$v is the 
local line width.
For these simulations, we generate synthetic line profiles of 
$^{13}$CO J=1-0 emission, including
excitation calculations up to J=5.  
Constant values of $^{13}$CO abundance relative 
to H$_2$ (1.25 $\times$ 10$^{-6}$) and a
kinetic temperature (20 K) are assumed. A pixel scale
of (20 pc/128) = 0.156 pc gives the H$_2$ column density within a cell 
at position (x,y,z) as 
$$ N_{H_2}(x,y,z) = 3.08{\times}10^{18}\; n(x,y,z)\; (20/128) \;\;\; cm^{-2}
\eqno(2.7) $$ 
The local
line width is determined from the quadrature sum of the thermal
velocity for the molecule, c$_{th}$, and the gradient
of the velocity field at that position 
$$ {\Delta}v(x,y,z) = \sqrt{c_{th}^2+
( { {\partial v(x,y,z)} \over {\partial z}} {\delta}z)^2} \eqno(2.8)   $$
Figure~4 shows the variations of excitation temperature and line 
center opacity  with density and  N$_{H_2}$X/${\Delta}$v for the assumed
kinetic temperature of 20 K.
Once the excitation temperature and line center opacity are 
determined for each point in the cloud, the emergent intensity at
velocity, $v$, 
is calculated for each position (x,y) 
$$ T(x,y,v) = { {c^2}\over{2k\nu^2}} \sum_{z=0}^{L_z} (B_\nu(T_{ex})-B_\nu(2.7))(1-e^{ -{\tau_\circ}
\phi(v(x,y,z),{\Delta}v)} )
e^{ -{\Upsilon}(x,y,z,v)} \eqno(2.9) $$
where B$_{\nu}$(T) is the Planck function, $\nu$ is the line frequency, 
$\phi$(v(x,y,z),${\Delta}$v) is the Gaussian profile function normalized 
to unity and 
$$ \Upsilon(x,y,z,v) = \sum_{\zeta=0}^z \tau(x,y,\zeta)\phi(v(x,y,\zeta,{\Delta}v)) \eqno(2.10) $$
is the total foreground opacity to a physical depth z.
The model cloud is placed at a distance of 1 kpc and is observed 
with a top-hat telescope beam response which covers 1 pixel or an 
effective angular resolution of 32.4". 
The spectral resolution of the observation is 0.05 kms$^{-1}$.
Examples of the synthetic observations can be found in Brunt (1999).

\section{Analysis}

The formal goal of PCA is to determine the set of orthogonal axes
such that the data, when projected onto these axes, maximizes the 
variance.  In practice, 
for spectroscopic 
imaging observations, it effectively identifies differences
in line profiles as these contribute to the variance of the data cube.
Such line profile differences can arise from dynamical processes within the gas
and can be 
distinguished from 
the instrumental noise of the data.  Therefore, PCA provides a powerful
method to identify kinematic variability within the target object.
In this section, we  describe the PCA technique and subsequent
structural analysis to
extract the observationally accessible
exponent $\alpha$ from spectroscopic imaging observations. 

\subsection{Principal Component Analysis Method}

A spectroscopic imaging observation is comprised of an ensemble of
$n=n_x{\times}n_y$ spectra each with p spectroscopic channels.
We write the resultant data cube as $T(x_i,y_i,v_j)=T(r_i,v_j)=T_{ij}$
where $r_{i}=(x_{i},y_{i})$
denotes the spatial coordinate of the $i$th spectrum.
The covariance matrix S$_{jk}$ is
$$ S_{jk} = {{1}\over{n}} \sum_{i=1}^{n} T_{ij}T_{ik} \eqno(3.1)$$
The set of 
eigenvectors, $u_{lj}$ and eigenvalues are
 determined from the solution of the eigenvalue equation for 
the covariance matrix, 
$$  Su = {\lambda} u \eqno(3.2) $$
The eigenvalue, $\lambda_l$, corresponds to the amount of variance projected 
onto its corresponding eigenvector, $u_l$. Therefore, the eigenvalues
(and corresponding eigenvectors) are reordered from largest to smallest.
In this implementation of PCA, we do not explicitly subtract the mean 
value, $<T_j>=\sum T_{ij}/n$ from each channel image. 
 These are effectively removed within the 
first, ($l=1$), principal component.
Note that no {\it spatial} information is
contained within the eigenvectors, since the ordering of the
spatial part of $T_{ij}$ is arbitrary.
The eigenvectors $u_{lj}$ are purely spectroscopic 
and trace the (ordered) sources of variance in the 
ensemble of observed line profiles. Also, the
eigenvectors are orthogonal, which ensures that the
information contained within each component is independent
information.   Similar decompositions can be obtained by other orthogonal
function
sets such as 
wavelets or spherical harmonics.
The advantage of PCA over these other basis sets is that the 
orthogonal vectors
are determined by the data and not predetermined.

To spatially isolate the sources of variance contained with the $l^{th}$
component,
an eigenimage, $I^l(r_i)$, is constructed from the projected values of the 
data, $T_{ij}$,  
onto the eigenvector, $u_{lj}$,
$$ I^{l}(r_{i}) = \sum_{j=1}^{p} T_{ij}u_{lj}  \eqno(3.3) $$
It is equivalent  to a velocity-integration of T(x,y,v)
that is weighted at each channel by the value of the appropriate
eigenvector.  
Since PCA is a linear decomposition, the random noise of the input 
data can readily be propagated.  The variance of projected values
due to noise is 
$$\sigma_l(r_i)^2 = \sum_{j=1}^{p} \left[{{\partial I^l(r_i)}
\over {\partial T_{ij}}}\right]^2 
                    \sigma(T_{ij})^2 = 
\sum_{j=1}^{p} u_{lj}^2 \sigma(T_{ij})^2 \eqno(3.4) $$
Assuming, that $\sigma(T_{ij})$ is 
constant over the bandpass of the 
spectrum and recalling that
 u is orthonormal ($\sum u_{lj}^2=1$), 
the expression reduces to 
$$\sigma_l(r_i) = \sigma(T_{ij}) \eqno(3.5) $$
Thus, the variance of values projected onto the $l^{th}$ component
is equal to variance of the input data.  More importantly, one
can readily distinguish values within the eigenimage from
those due to instrumental noise.

\subsection{Characteristic Scale Measurements}

For a given component, $l$, the eigenvectors and eigenimages
describe the velocities and positions at which the measured 
line profiles are different with respect to the noise level.
Characteristic velocity differences and the spatial scales 
over which these differences occur are derived from e-folding
lengths of the normalized autocorrelation functions (ACFs) of the 
eigenvectors and eigenimages respectively (Heyer \& Schloerb 1997).
The raw autocorrelation functions, $C_V^l(dv)$, $C_I^l(\tau)$, 
are calculated directly from the 
eigenvectors and eigenimages respectively,
$$ C_{V}^{l}(dv) = < (u^{l}(v) u^{l}(v+dv)> \eqno(3.6a) $$
$$ C_{I}^{l}(\tau) = < (I^{l}(r) I^{l}(r+\tau)> \eqno(3.6b) $$
However, in this form, the 2 dimensional autocorrelation function 
contains an additive component, $C_N(\tau)$, due to instrumental noise.
In Section 3.3, we demonstrate that the noise subtracted autocorrelation
function, $C_{I0}^l(\tau) = C_I^l(\tau) - C_N(\tau)$
where $C_N(\tau)$ is the autocorrelation function of noise values of the data.

The characteristic velocity scale, ${\delta}v_l$, is determined
from the velocity lag at which $C_V^l({\delta}v_l)/C_V^l(0) = e^{-1}$.
For the 2 dimensional ACF of the eigenimage, one must additionally 
correct for the 
effects of finite resolution observations upon the zero lag (see Section 3.4).
We determine the spatial
correlation lengths along the cardinal directions, ${\delta}x_l, {\delta}y_l$,
 from the noise corrected
ACF,
$$ { {C_{I0}^l({\delta}x_l)}\over{C_{I0}^l(0)} } =e^{-1} \eqno(3.8a) $$
$$ { {C_{I0}^l({\delta}y_l)}\over{C_{I0}^l(0)} } =e^{-1} \eqno(3.8b) $$
and assign the {\it biased} characteristic spatial scale, $L_{lB}$ to the
quadrature sum,
$$ L_{lB} = \sqrt{{\delta}x_l^2+{\delta}y_l^2} \eqno(3.9) $$
As discussed in Section 3.4, the true characteristic scale, $L_l$, 
is derived from the $L_{lB}$ with subpixel corrections to account for 
finite resolution of the observations.
Since the eigenimages tend to be isotropic, the ACF profiles along 
the cardinal directions are a reasonable approximation to the variation 
of the ACF along any arbitrary angle.
The characteristic velocity and spatial scales are statistically well defined 
with respect to the noise as these represent mean values over the full spectrum
and field respectively.   A measurement error, $\sigma_{{\delta}v}$, to the
value ${\delta}v_l$ is given by one half the velocity resolution
of the spectrometer.  For the spatial scale, $\sigma_L$ is determined from
the quadrature sum of the spatial resolution 
and the degree of 
anisotropy, $|{\delta}x_l-{\delta}y_l|$, in the spatial ACF.
The PCA velocity statistic, $\alpha$, is then 
obtained from the set of ${\delta}v,L$ pairs which are larger than the 
respective spectral and pixel resolution limits and form the power law 
relationship
$$ {\delta}v \propto L^{\alpha} $$
where the unsubscripted quantities refer to the ensemble
(${\delta}v_l, L_{l}$). Note that in the retrieval
of each ${\delta}v-L$ pair, PCA incorporates
information from the entire data cube such that each pair 
is well defined. 

\subsection{Instrumental Noise Effects}

All real observations contain a noise contribution.  For any 
analysis, it is critical to evaluate how the noise may 
affect the result.
For the ACF of the eigenvector, 
the characteristic velocity scale measurements are largely unaffected 
by 
instrumental noise due to the limited number of channels (p$<$128). 
However, for large, 2 dimensional images, the contributions to the ACF due to 
noise can be significant and must be removed.
The observed eigenimage is the sum of a noiseless eigenimage, $I_0^l(r)$,
and a noise contribution, $N(r)$,
$$ I^{l}(r) = I^{l}_{0}(r) + N(r) \eqno(3.10) $$
The unnormalized spatial ACF,  
($C_{I}^{l}(\tau)$) is 
$$ C_{I}^{l}(\tau) = < (I^{l}_{0}(r)+N(r))(I^{l}_{0}(r+\tau)+N(r+\tau))> \eqno(3.11) $$
$$ C_{I}^{l}(\tau)=C^{l}_{I0}(\tau) + C^{l}_{N}(\tau)+<I^{l}_{0}(r)N(r+\tau) + I^{l}_{0}(r+\tau)N(r) > \eqno(3.12) $$
where C$^{l}_{I0}$ is the ACF of the noiseless eigenimage
and C$^{l}_{N}$ is the ACF of the noise.
The term in brackets measures the correlation between signal
and noise. We assume that the signal and noise are
uncorrelated and treat this term as zero.  Therefore, 
$$ C^{l}_{I0}(\tau) = C^l_{I}(\tau) - C_{N}^l(\tau) \eqno(3.13) $$
For Gaussian noise, 
\[
C_{N}^{l}(\tau) = \left \{
\begin{array}{ll}
  \sigma^{2}_{N} & \mbox{ for $\tau  =  0$} \\
  0 & \mbox{ for $\tau  \neq  0$}
\end{array}
\right.
\]
We exploit the noise propagation properties of PCA and obtain 
$\sigma^{2}_{N}$ from several high $l$ eigenimages,
which are free of signal. 
A more general procedure can be followed to remove the noise
contributions to the autocorrelation function.   The noise ACF,
$C^l_{N}(\tau)$, can be calculated directly from either channel
images within the data cube or high $l$ 
eigenimages which contain no signal.   The noise contribution 
is removed at each value of $\tau$.  This method works well
for non-Gaussian noise fields such as those obtained by 
on the fly mapping or reference sharing observations (Brunt \& Heyer 2000).

The noise subtracted ACF, $C^l_{I0}$ is well determined even 
in the presence of relatively high noise levels. 
If this noise subtraction is not carried out, then 
C$^{l}_{I}$ contains a noise spike at the origin,
such that the spatial correlation lengths are underestimated.
Since the noise contribution relative to the signal is larger for higher $l$ 
eigenimages, this means that the scale lengths are
underestimated in such a way as to lessen the slope $\alpha$
of the of the log(${\delta}v$)-log($L$) relationship.
HS97 did not make this correction for the noise contribution.
Therefore, the derived values of 
$\alpha$ obtained by HS97
should be considered as 
lower limits. 

\subsection{Correction for Finite Resolution Effects}

It is important that the analysis results are independent 
of the resolution at which the
observations are obtained. In this Section we discuss the
systematic effects within the ACFs due to finite
resolution.  We demonstrate in Section 4.4 that
the spectroscopic ACFs and the derived velocity
scales 
are not compromised by changing the
spectroscopic resolution at which the observations are
obtained.  However, the eigenimage ACFs {\it are} resolution-dependent
and require a sub-pixel correction to
ensure that the spatial scale measurements are not
systematically compromised.
In brief, for finite resolution observations, 
the unnormalized ACF at zero lag, $C^l_{I0}(0)$, is underestimated.
However, by assuming, or estimating, a shape to the ACF in order to 
extrapolate to zero lag,
one can gauge the amount that the scale length is overestimated 
and provide a correction to the derived biased scale lengths, $L_{lB}$.
The alternative option is 
to fit a functional form with a ``scale'' parameter to
the ACF. 
We have found that making small
corrections to scales obtained from a biased ACF provides the
more stable measurement.

A detailed derivation of the corrections for various functional 
forms of the ACF is given by Brunt (1999).  In brief, the 
measured 
value of $C_{I0}(0)$ which is used to normalize the ACF is 
$$ C_{I0}(0) \approx C(\epsilon L_{pix}) \eqno(3.14) $$
where $C(\tau)$ is the true ACF, 
$L_{pix}$ is the pixel size, and $\epsilon < 1$, 
is a number that depends on the shape of the ACF and the
shape of the telescope beam.  For
eigenimages derived from observations with a top hat beam profile,
$\epsilon \approx 0.5$. This underestimation with respect to
an ideal ACF applies to 
the full resolution (128$^{2}$) eigenimages, since a 
simulation or observation at any finite resolution necessarily involves
omission of variance (i.e. contributions to $C_{I0}$)
that would be present in a higher resolution measurement. 
Fortunately, the degree of underestimation and the
associated scale corrections for the
128$^{2}$ eigenimages are small.

The overestimated scale measurement $L_{B}$ obtained from an
underestimated ACF, $C_{I0}$,
satisfies
$$ {{C(L_{B})}\over{C_{I0}}} = exp(-1) \eqno(3.15) $$
For a functional form of the true
ACF, $C$
$$ C(\tau) = exp(-(\tau/L)^{\kappa})  \eqno(3.16) $$
where $L$ is the true scale length of the ACF and $\kappa$
parameterizes the shape of the ACF, 
the biased scale length $L_{B}$ is related to the
true scale length $L$ via
$$ L \approx (L_{B}^{\kappa} - (\epsilon L_{pix})^{\kappa})^{1/\kappa} \eqno(3.17) $$
with $\epsilon \approx 0.5$.
For $\kappa=1$ (an exponential ACF), 
the correction to all measured scales  is 
$-{\epsilon}L_{pix} \approx -L_{pix}/2$.
For a
Gaussian beam 
$-{\epsilon}L_{pix} \approx -0.75L_{pix}$ (Brunt 1999).
The form of the ACFs for each observation is estimated
from a few low $l$ eigenimages. 
The corrections are
not sensitively dependent on small errors in estimating
$\kappa$.  
The application of these corrections to the measured scale lengths 
ensures consistency of the results for varying resolutions.

\section{Results}

Initially, the 
basic calibration between the observed exponent $\alpha$ and the
intrinsic field statistics, parameterized by $\beta$,
is established for uniform density fields. 
Subsequently, we evaluate the ${\delta}v-L$ relationship 
with respect to the uniform density results
using more realistic distributions of interstellar material and 
observational conditions 
including
instrumental noise
and varying resolution.

\subsection{Uniform Density Results}

Ten realizations of the velocity field are created for each 
value of $\beta$ ranging from 1.0 to 4.0
according to the prescription given in Section 2.1. The real
space statistics of these fields are checked and reported
in Section 2.2.  Synthetic line profiles of $^{13}$CO J=1-0 emission
are calculated at each position, (x,y),
 assuming a 
uniform
density (n$_{H_{2}}$ = 100 cm$^{-3}$) and uniform temperature (T$_{k}$ = 20K)
via the LVG radiative transfer method described in Section 2.5.
The optical depths in these observations are small and
the excitation is subthermal and uniform throughout the field 
which corresponds to an
almost perfectly-sampled velocity field. 
The resultant data cubes are decomposed using PCA and velocity 
and spatial scales are estimated from the eigenvectors and 
eigenimages respectively.  These scales define a relationship between
the magnitude of velocity differences and the spatial scales at which 
these differences occur.   The relationship is fit to a power law
parameterized by the exponent $\alpha$.

Figure~5 shows ${\delta}v-L$
relationships for 6 velocity fields with varying values of $\beta$. 
The dotted lines mark the spectroscopic
and spatial pixel sizes. Uncorrected spatial scales
$L_{B}$ and spectroscopic scales
${\delta}v$ are restricted to $>2$ pixels and
$>1$ spectroscopic channel respectively (see Section 3.4).
The lower $\beta$ simulations
are characterized by a fewer number of retrieved ${\delta}v-L$
pairs above the resolution limits since most of the variance is 
generated at small scales. 

In Table~1  and Figure~6 we show the relationship
between the mean value of $\alpha$ determined from the ensemble 
of realizations  and the input $\beta$. 
The plotted error bars
reflect the standard deviation of values from this mean.
These results show that there is a monotonic
relationship between the observed and intrinsic exponents for 
$\beta < 3$ and the method, as described in Section 3, is 
sensitive to velocity fields with varying statistics.
A least squares fit to the $\alpha,\beta$ values for 
$1 \le \beta \le 2.5$ gives
$$ { \alpha = (0.33 \pm 0.04)\beta - 0.05 \pm 0.08 } \eqno(4.1) $$
This was obtained with $\alpha$ as the dependent variable.
For values
of $\beta >3$, the derived values of $\alpha$ show a similar transition
 to smoothness that was seen
in the intrinsic $\gamma$ measurements described in Section 2.2.
Equation 4.1 provides the basic calibration between an observable
measure of kinematic variability within a data cube and the intrinsic velocity 
field statistics.  The relationship between $\alpha$ and $\beta$
remains an empirical measurement.  It surely depends upon the 
dimensionality of the fields since a recent calibration of this 
technique to 2 dimensional fields reveals a modified relationship,
$\alpha_{2D}\approx\beta_{2D}/4$.  Given the 
extreme complexity of transforming a velocity field onto a spectroscopic 
axis and the identification of velocity gradients by line profile differences,
an analytical description of equation 4.1 is not readily determined.

\subsection{Lognormal Density Field Results}

The results of the previous Section are an important
first step in gaining insight into how PCA retrieves
the intrinsic velocity field information.
However, a uniform density field
is clearly unphysical.  To evaluate the method with a more 
realistic density distribution, 
lognormal density fields (see Section 2.3)
with mean values
of 10$^2$, 10$^3$, and 10$^4$ \cc~ and velocity fields
with $\beta=1,2,3$ are observed.  The increase of the mean
density corresponds to an increase in gas column density and 
line center optical depth within a given cell.  
 The primary effect of
inhomogeneous
density fields is to mask the velocity field from the 
observer as the low density regions do not sufficiently 
excite the molecular gas tracer into emission. 
Figure~7 shows the ${\delta}v-L$ relationships obtained
by PCA from these observations for these varying conditions. 
The results obtained from
the uniform density observations with the identical velocity
fields are included for comparison. The resultant $\alpha$
measurements are tabulated in Table~2.

For the ${\beta} \ge 2$, the derived values of $\alpha$ are
similar to the uniform density exponents although there is
a constant offset in the logarithm.  In addition, the
number of significant components at the largest spatial scales
increases with increasing density and optical depth.  This is 
surely due to radiative trapping within a given cell 
which maintains an elevated
excitation temperature in low density regions in which the trace
molecule  would otherwise be insufficiently excited.
This suggests that a 
molecular gas tracer with moderate to high optical depths, 
such as $^{12}$CO, provides a more
complete sampling of the velocity field.
The
$\beta=1$ observations tend to overestimate $\alpha$
relative to the uniform density exponent.  In this case, 
the presence of high frequency velocity fluctuations
places many cells along a given line of sight
at a common velocity (see Figure 1).  Therefore, for optical depths 
larger than unity,
the back side of the cloud is hidden from the observer and the field 
is undersampled.  
From these results, we conclude that the velocity field statistics
with inhomogeneous density fields can be reliably recovered 
for $\beta \ge 2$ under a wide range of optical depths but are
slightly overestimated for more shallow energy spectra $\beta \sim 1$.

\subsection{Effects of Instrumental Noise}

Any real observation of the ISM necessarily contains instrumental
noise.  In this Section, we evaluate the effects of instrumental noise upon 
the analysis by 
adding a noise field, N(r,v), to the  simulated observations
of the  lognormal density field ($<n_{H_{2}}> = 10^{4}$ cm$^{-3}$)
with values of $\beta = 1,2,3$.  The input noise is
Gaussian with variance $\sigma^{2}_{N}$. 
We define $\zeta$ 
as a gauge of the
signal to noise of a data cube, 
$$ \zeta = \sqrt{{{\sigma^{2} - 
\sigma^{2}_{N}}\over{\sigma^{2}_{N}}}} =  
{ {\sigma_{0}}\over{\sigma_{N}}} \eqno(4.2) $$
where $\sigma^{2} = \sigma^{2}_{0} + \sigma^{2}_{N}$ is
the variance of the data cube with noise 
and $\sigma^{2}_{0}$ is the variance of the same in
a noiseless observation.  Since PCA
generally consolidates all spectroscopic channels with signal 
within the first principal component and noise in the higher components,
$\zeta$ can be calculated from the $l=1$ and $l=p$ eigenvalues, 
$$ \zeta = \sqrt{{{\lambda_{1} - \lambda_{p}}\over{\lambda_{p}}}} \eqno(4.3) $$
 
In Figure~8, we show the ${\delta}v-L$ relationships 
derived from simulations with varying signal to noise
($\zeta = 4.0,2.0,1.0$).  Also shown are the 
relationships derived with $\zeta=\infty$ (dotted line).
The resultant $\alpha$ measurements are shown in
Table~3.
The primary effect of noise 
is to reduce the number of significant
components that are detected above the resolution
limits as the variance generated by the velocity field 
which is contained in the higher components are overwhelmed
by the variance due to noise.
This 
tends to induce systematic effects at large $\beta$
rather than an increased scatter in the retrieved
${\delta}v-L$ pairs.  For most spectral line
imaging observations,
$\zeta > 4$ for which the results are largely unaffected 
by noise (Brunt \& Heyer 2000).

\subsection{Resolution Effects}

In Section
3.6, we described the necessary corrections to the retrieved 
spatial scales which minimize the effects of
finite resolution observations.  In this section, we demonstrate the 
need and utility for such corrections using the simulated 
observations with varying resolution.
The original simulated observations are comprised of
128 $\times$ 128 spatial pixels.  
We degrade the resolution of the 
simulated 
observation ($\beta=2$ and lognormal density field with $<n(H_2)>=10^4$
 cm$^{-3}$)
by spatial block-averaging.   
This is a reduction in the
number of pixels and is equivalent to observing the simulation
at a further distance with a top hat beam.
We
define $\Lambda_I$ as the number of resolution elements across
the image.
Block averaged versions of this simulated observation are 
calculated for $\Lambda_{I} = 64,32,16$. Further
reduction in resolution results in only one detected measurement
above the resolution limits. The measurements obtained from
these observations, for both uncorrected scales and 
scales corrected for an exponential ACF ($\kappa=1$)
are shown in Figure~9.
These plots
demonstrate the saturation of retrieved pairs at the limit
$L = (1 - exp(-1))L_{pix}$, expected for a linear interpolation
between unity and zero in the normalized ACF.
This feature is 
characteristic of uncorrelated noise with
a true scale length of zero. The uncorrected
measurements clearly show steepening due to a uniform scale
overestimation as the resolution limit, $L_{pix}$, 
is approached.  It is this steepening that compromises the estimation of
$\alpha$. 
When the corrections are applied to the retrieved spatial scales, the
derived values of $\alpha$ are consistent between the varying resolutions.
Note that if no corrections are made, then the measured value of 
$\alpha$ represents
an upper limit, obtained with the {\it assumption} that there
is no sub-pixel inhomogeneity. Measurements on real data will always
contain this ambiguity, which can only be resolved
by increasing the resolution.

It was stated in Section 3.4 that the velocity scales ${\delta}v$ are
unaffected by changing the spectroscopic
resolution. To demonstrate this statement, 
we coarse-grained the velocity axis of the same observation.
In analogy to the above discussion of spatial scales, we
define $\Lambda_{v}$ as the ratio of the 
total range in velocity spanned by
the spectrometer and the velocity resolution.
For the original simulated observation,
$\Lambda_{v}=128$.
The spatial resolution is fixed at $\Lambda_{I}=128$.
The results of the PCA retrievals
for these fields are shown in Figure~10.  Also shown are scale
measurements for 
a coarse-grained field with 
$\Lambda_{I}=64$, and $\Lambda_{v}=32$.  
The spatial scales have been corrected for finite spatial resolution.
No corrections have been applied to the retrieved velocity scales. 
These results shows that
the velocity scale measurements with degraded spectral resolution
are equivalent to those derived from the full resolution observations.

\section{Summary }

We have investigated the ability of PCA to recover 
intrinsic turbulent velocity field statistics
from spectral line imaging observations of the molecular
ISM, and established a preliminary calibration of the
method. The calibration, subject to the limitations of
the modeling, identifies a monotonic relationship
between the intrinsic $\beta$ values (or equivalently, $\gamma$) 
and observed 
exponents ($\alpha$) that characterize the velocity field statistics.
 
In addition to the simplistic nature of the input density
and velocity fields, including the assumption of
statistical isotropy and the periodicity feature,
the major limitations are the
lack of mutual consistency between density and velocity,
and the assumption of local excitation in the radiative
transfer calculations. However, the simulated lognormal observations
are in good visual correspondence with
real molecular ISM observations. The retrievals
from the low density observations are in
close agreement with those from the high
density observations. This suggests that the accuracy
of the method depends upon the degree to which one
samples the velocity field,
rather than the details
of the radiative transfer. 
For molecular clouds in particular, highly saturated
observations may be more reliable (due to more
homogeneous sampling and a greater
number of retrieved measurements) and this
suggests that $^{12}$CO is likely to provide the most
accurate values of $\alpha$. 
This method could also
be applied to HI 21cm emission which is excited over a
wide range of physical conditions.

The sampling effects due to noise, which uniformly masks
sources of variance, do not affect the fits for $\alpha$,
until very high noise levels are reached. As calibrated, the 
PCA method is
resolution-independent but limited 
in general by necessary assumptions of sub-pixel homogeneity
of the emission field. 
Independent of our modeled calibration, the robustness of
measured $\alpha$ values provide quantitative constraints 
to simulated observations of velocity and density fields 
from hydrodynamical calculations. We do stress, however,
that the calibration of $\alpha$ to $\beta$ achieved in this
paper is strictly only valid for the fBm model velocity
fields. More realistic turbulent fields may not adhere as closely
to the calibration, and in future we intend to include
supercomputer MHD simulations in the PCA calibration.

Empirical measurements of $\alpha$ from real data of course do not 
depend on the calibration to $\beta$ (or $\gamma$), and such 
measurements will be reported by Brunt \& Heyer (2000) for
an ensemble of fields taken from the FCRAO
Outer Galaxy Survey. Using the preliminary
calibration established here, we obtain
constraints on the intrinsic velocity field
statistics of the molecular ISM.

This work was supported by NSF grant AST 97-25951 to the Five College
Radio Astronomy Observatory. 

\clearpage
\section{References}
\begin{description}
\item[] Arquilla, R. \& Goldsmith, P.F. 1984, ApJ, 279, 664
\item[]Brunt, C.M., 1999, Ph.D dissertation, University of Massachusetts
\item[]Brunt, C.M. \& Heyer, M.H. 2000, in preparation
\item[] Carr, J.S. 1987, ApJ, 323, 170
\item[] Dickman, R.L. 19815, in Protostar and Planets II, eds. D.C. Black
\& M.S. Matthews, 
University of Arizona Press, p. 150
\item[] Falgarone, E., Phillips, T.G. \& Walker, C.K. 1991, ApJ, 378, 186 
\item[] Goldreich, P. \& Kwan, J. 1974, ApJ, 189, 441
\item[] Heyer, M.H. \& Schloerb, F.P., 1997, ApJ, 475, 173 (HS97)
\item[] Kleiner, S.C. \& Dickman, R.L. 1985, ApJ, 295, 466
\item[] Kolmogorov, A.N, 1941, Dokl. Akad. Nauk SSR, 30, 301
\item[] Larson., R.B., 1981, M.N.R.A.S, 194, 809
\item[] Leung, C.M. \& Lizst, H.S. 1976, ApJ, 208,  732
\item[] Miesch, M.S. \& Bally, J., 1994, ApJ, 429, 645
\item[] Myers, P.C. \& Gammie, C.F. 1999, ApJ, 522, L141
\item []Padoan, P., Jones, J.T. \& Nordlund, A.P., 1997, ApJ, 474, 730
\item []Passot, T. \& Vazquez-Semadeni, E., 1998, Phys. Rev. E, 58, 4501
\item []Pecknold, S., Lovejoy, S., Schertzer, D., Hooge, C. \& Malouin, J.F., 1993, in 
Cellular Automata : Prospects in Astrophysical Applications,
eds. Perdang, J.M. \& Lejeune, A., (World Scientific), p. 228
\item []Scalo, J.M., 1984, ApJ, 277, 556
\item []Scalo, J.M., 1990, in Physical Processes in Fragmentation and Star
Formation, eds. R. Capruzzo-Dolcetta, C. Chiosi, \& A. di Fazio,
(Dordrecht: Kluwer), p 151
\item []Schertzer, D. \& Lovejoy, S., 1987, J. Geophys. Res., 92, 9693 
\item []Scoville, N.Z., \& Solomon, P.M., 1974, ApJ, 187, L67.
\item []Solomon, P.M., Rivolo, A.R., Barret, J., \& Yahil, A. 
1987, ApJ, 319, 730
\item[] Stutzki, J. \& Gusten, R, 1990, ApJ, 356, 513
\item[] Stutzki, J., Bensch, F., Heithausen, A., Ossenkopf, V., \& Zielinsky, M.
 1998, AA, 336, 697
\item[] Voss, R., 1988, in The Science of Fractal Images, Eds. Peitgen, H.O., Saupe, D. (New York:Springer-Verlag) 
\item []Williams, J.P, de Geus, E.J. \& Blitz, L., 1994, ApJ, 428, 693
\item []Zuckerman, B. \& Evans, N.J. 1974, ApJ, 192, L149

\end{description}
\clearpage
\begin{figure}
\plotone{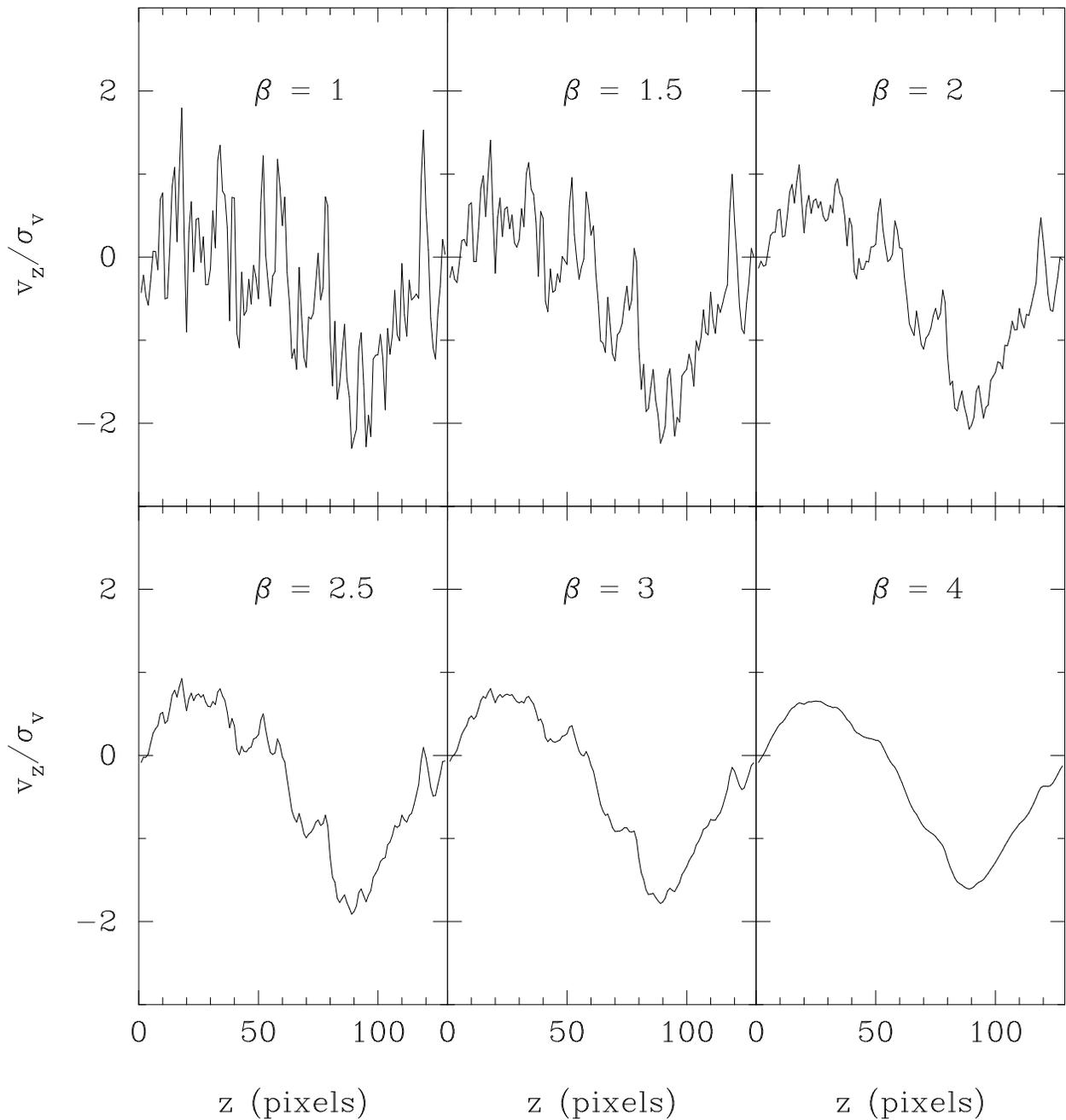}
\caption
{Examples of one-dimensional cuts through the 3 dimensional 
model velocity fields for $ 1 \le \beta \le 4$.}
\end{figure}

\begin{figure}
\plotone{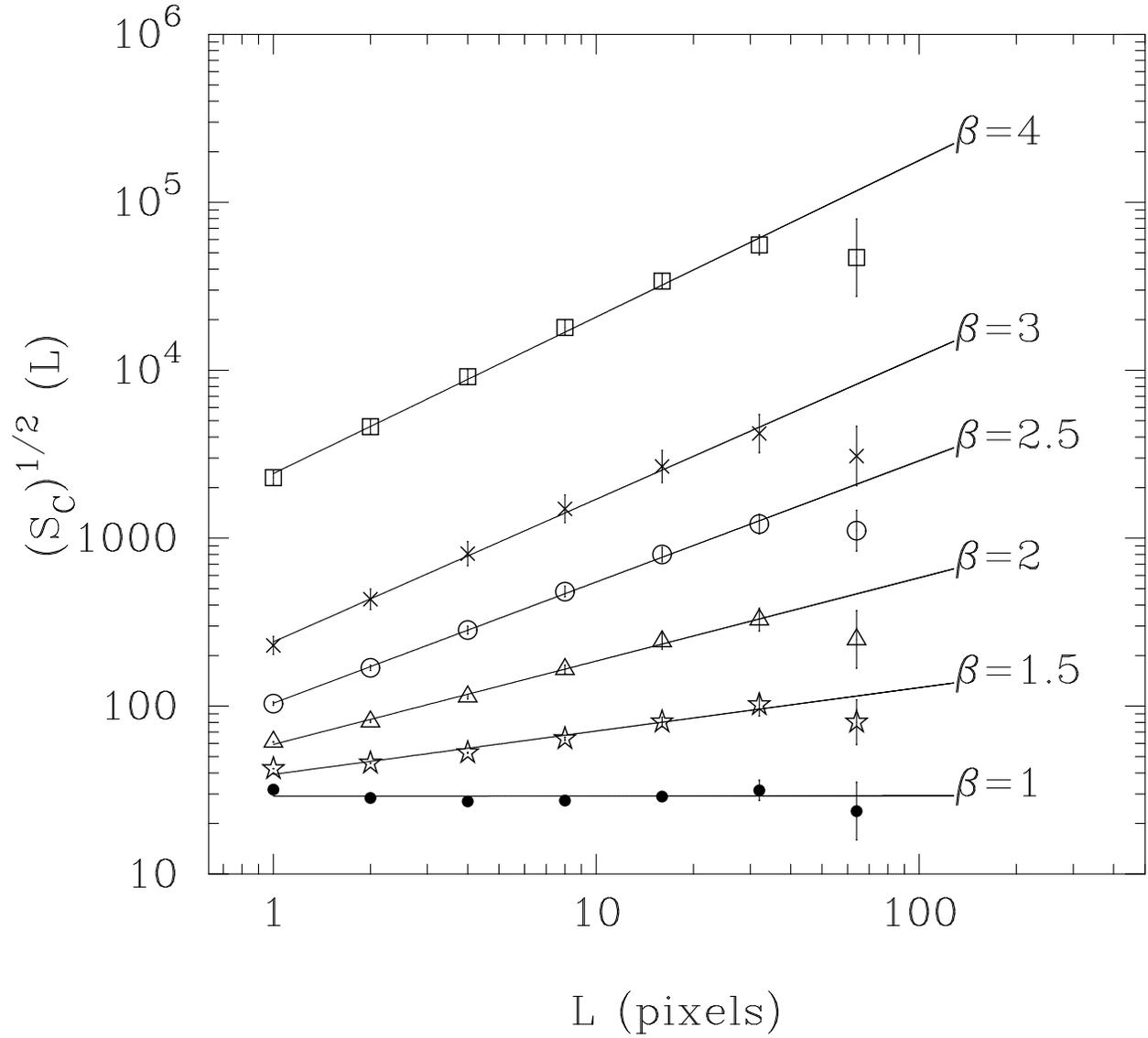}
\caption
{Ensemble coarse structure
function measurements of the model velocity fields.
Major tick marks correspond to powers of ten.}
\end{figure}

\begin{figure}
\plotone{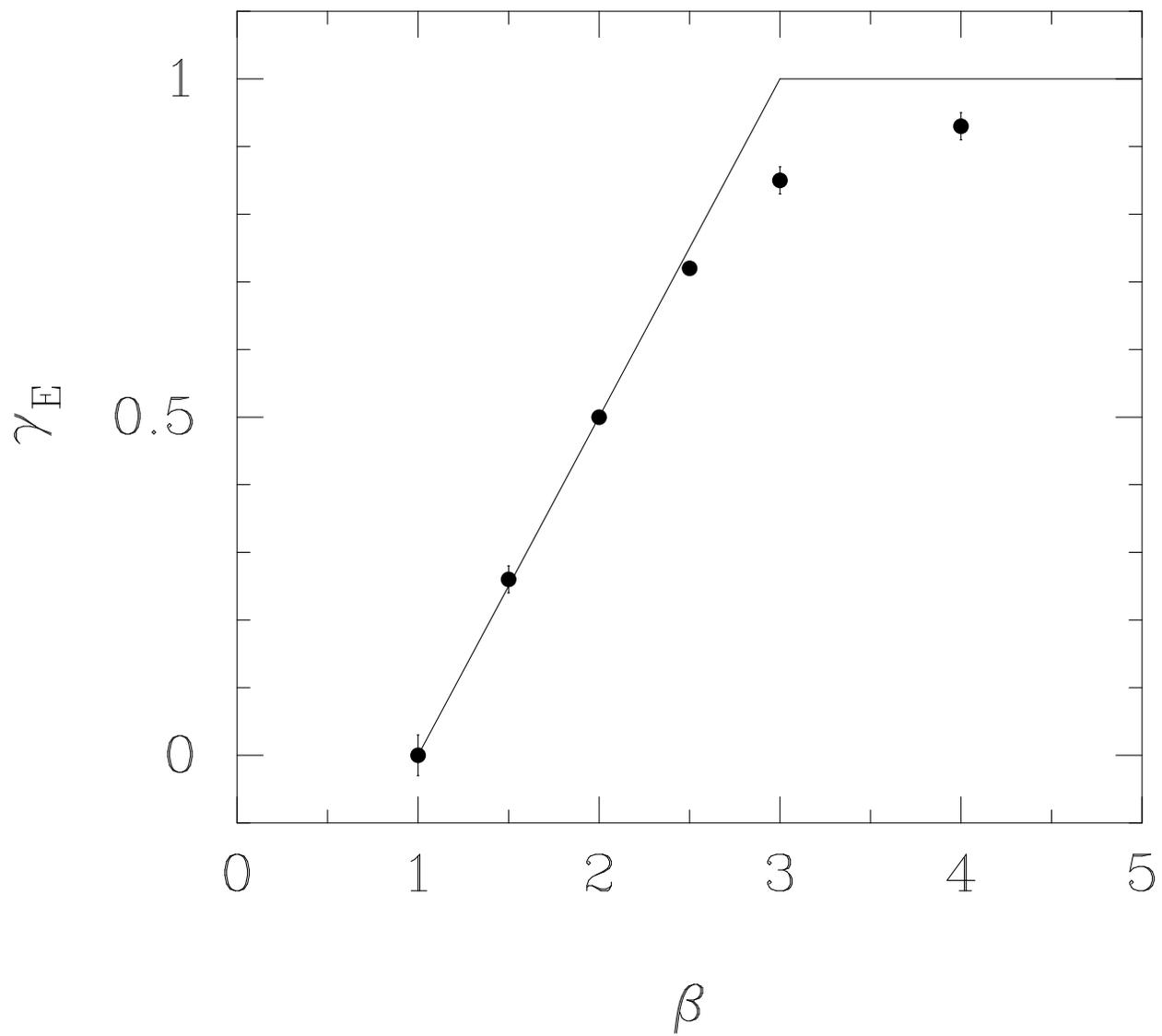}
\caption{Measured relationship between $\beta$ and $\gamma$
derived from the coarse structure function.}
\end{figure}

\begin{figure}
\plotone{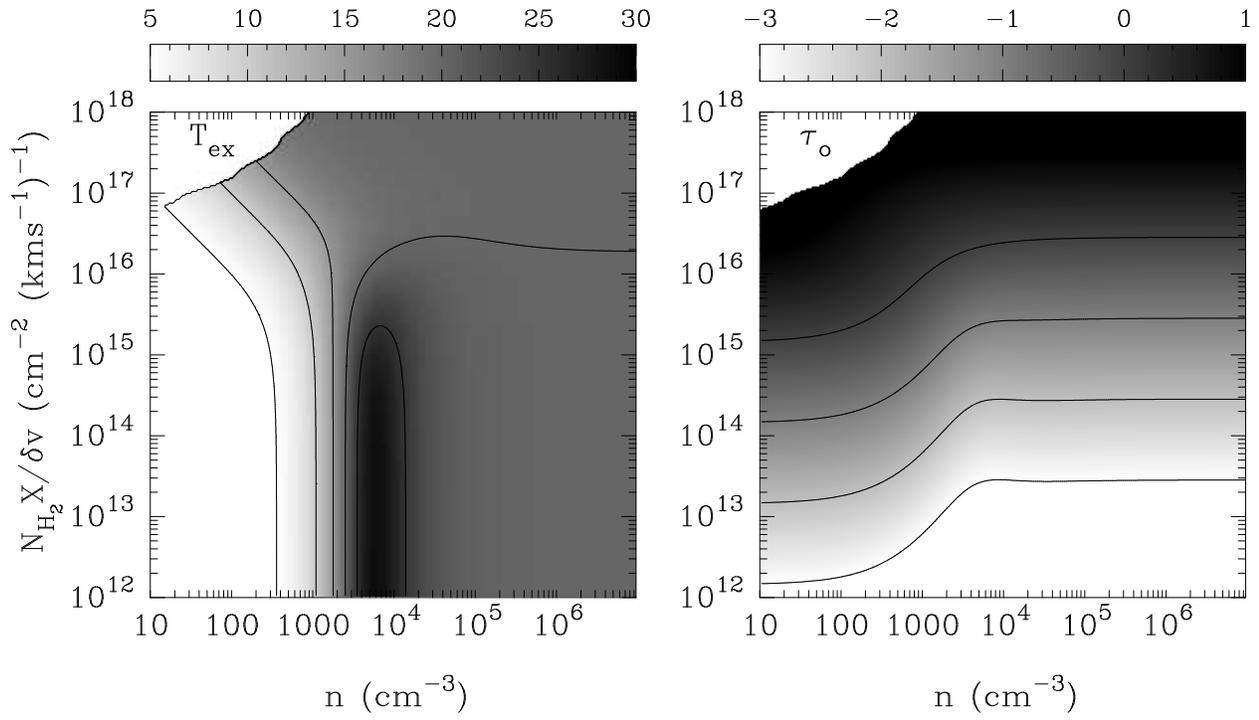}
\caption
{Dependence of $T_{ex}$ (left) and $\tau_{\circ}$ (right)
on the local (pixel) density, column density and velocity gradient.
The $\tau_{\circ}$ wedge is log$_{10}$arithmic.}
\end{figure}

\begin{figure}
\plotone{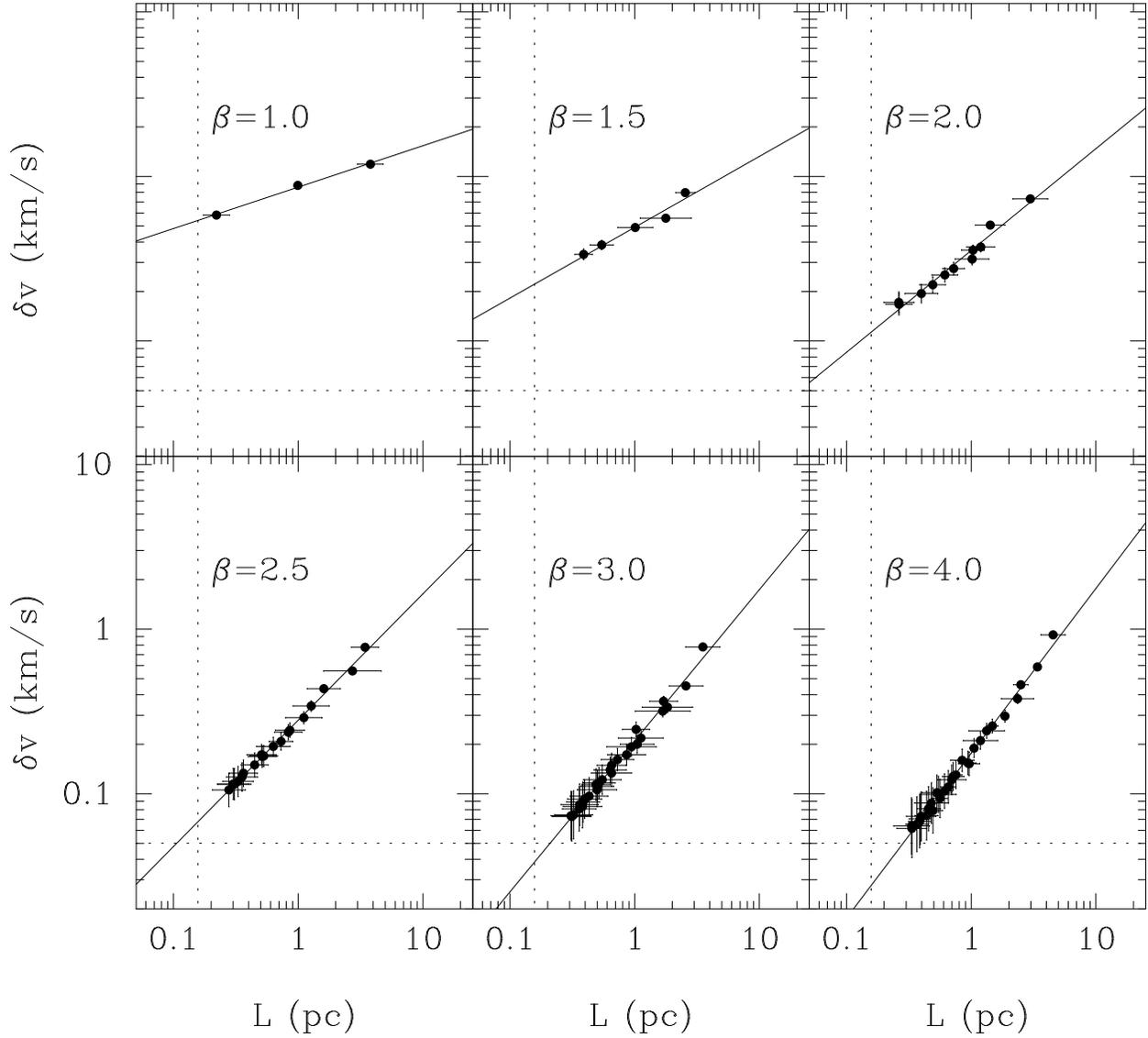}
\caption
{Representative results of the PCA retrieval for the uniform density
observations. The dotted lines mark the spatial pixel size
and spectroscopic channel width.}
\end{figure}

\begin{figure}
\plotone{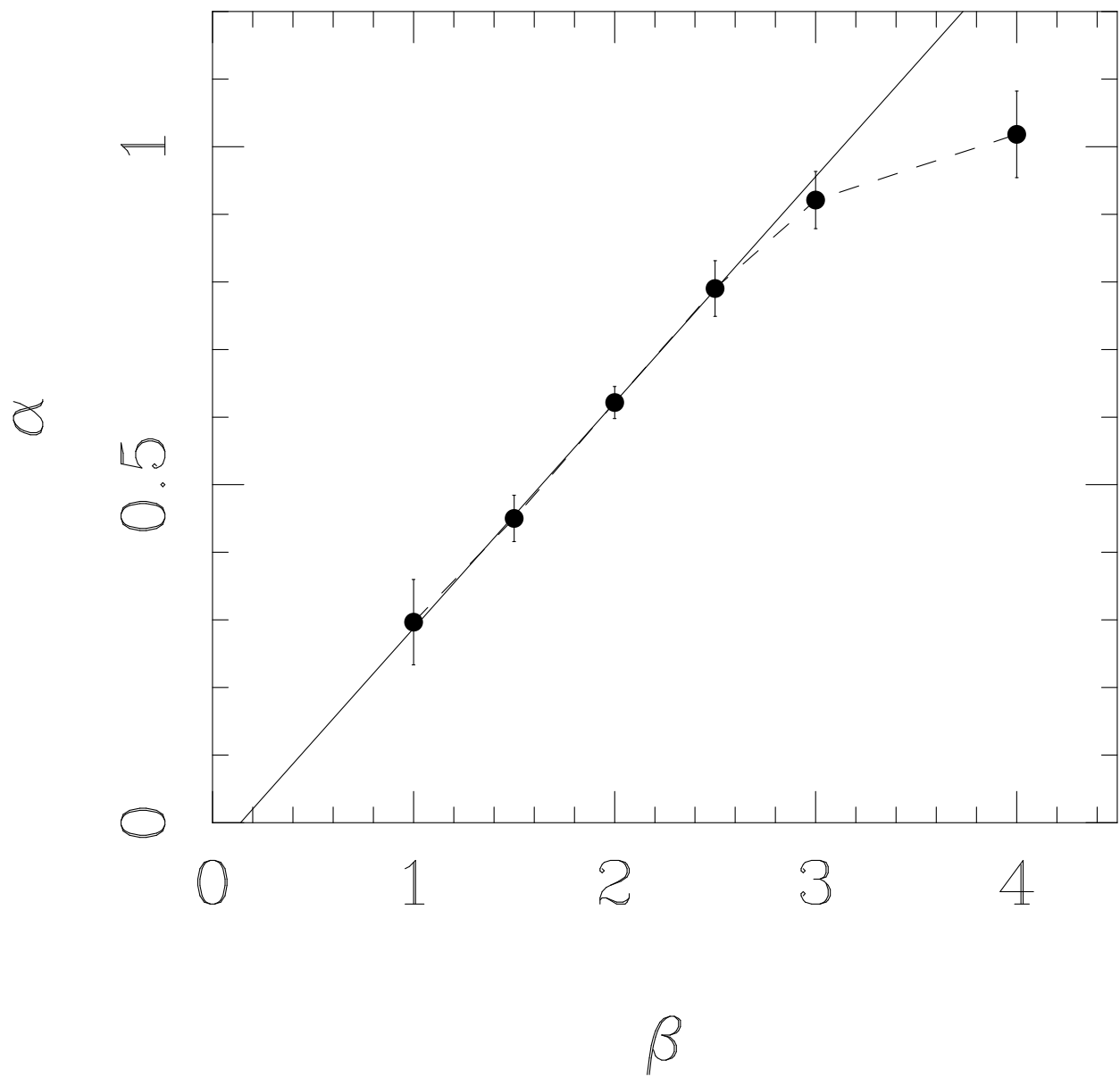}
\caption
{Basic calibration of the observable $\alpha$ exponent to the
intrinsic exponent $\beta$.}
\end{figure}

\begin{figure}
\plotone{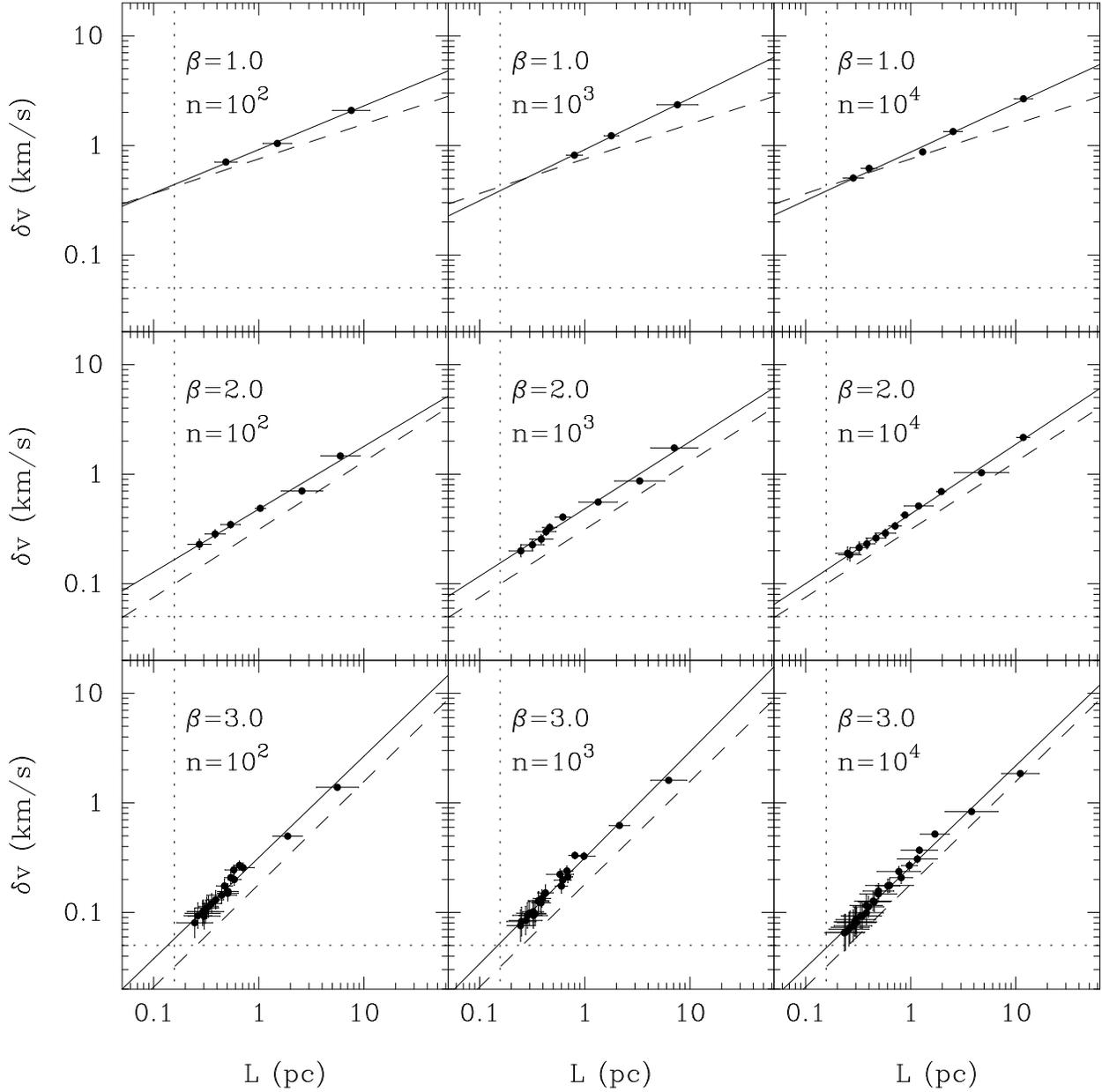}
\caption
{PCA measurements from the observation of the lognormal
density field. The dashed lines denote the fits obtained from
the observation of the same velocity fields with a uniform
density distribution. The solid lines are the fits to
${\delta}v$-L pairs.}
\end{figure}

\begin{figure}
\plotone{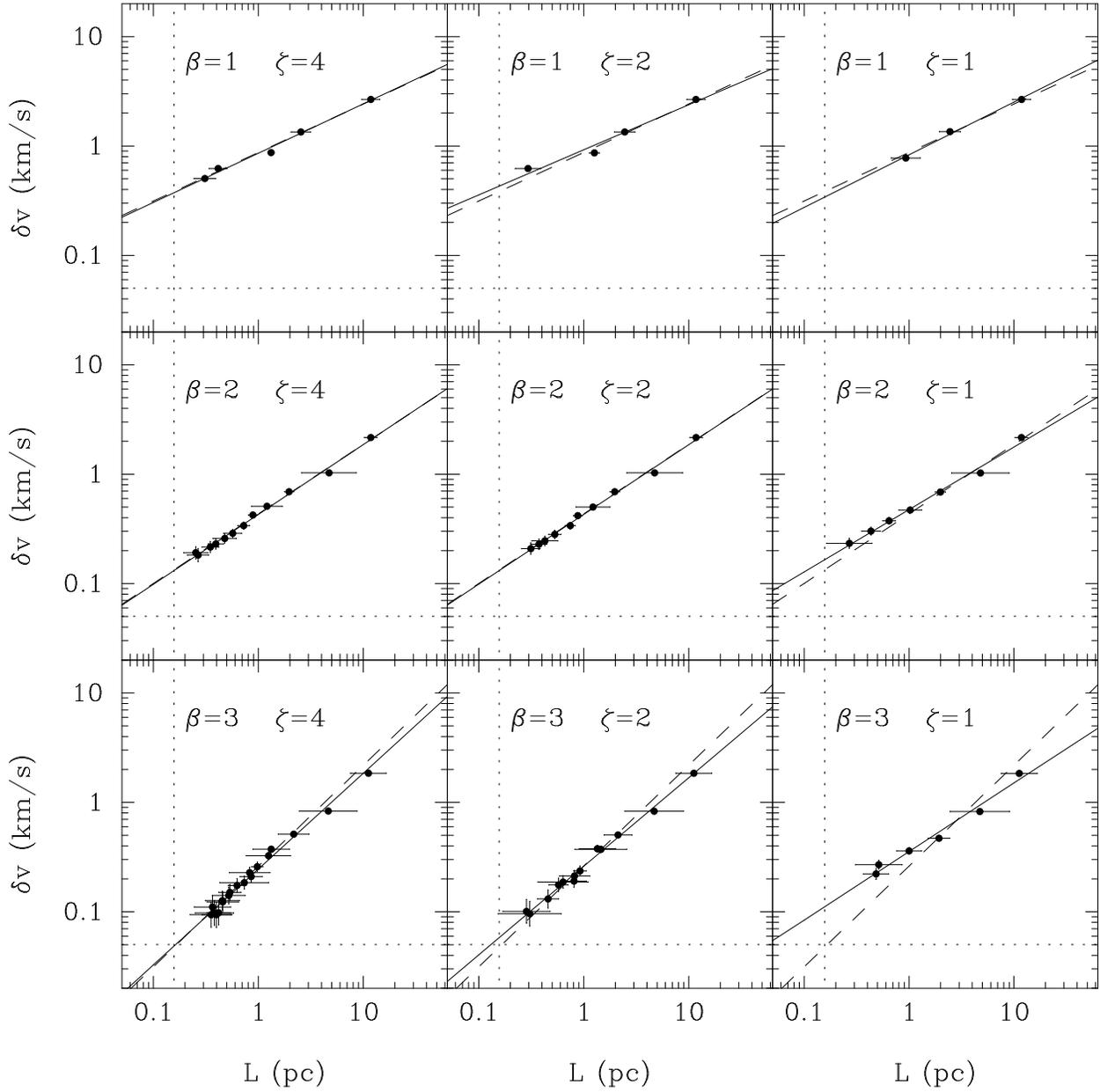}
\caption
{PCA measurements from observations of the lognormal
density field with varying degrees of gaussian distributed noise.}
\end{figure}

\begin{figure}
\plotone{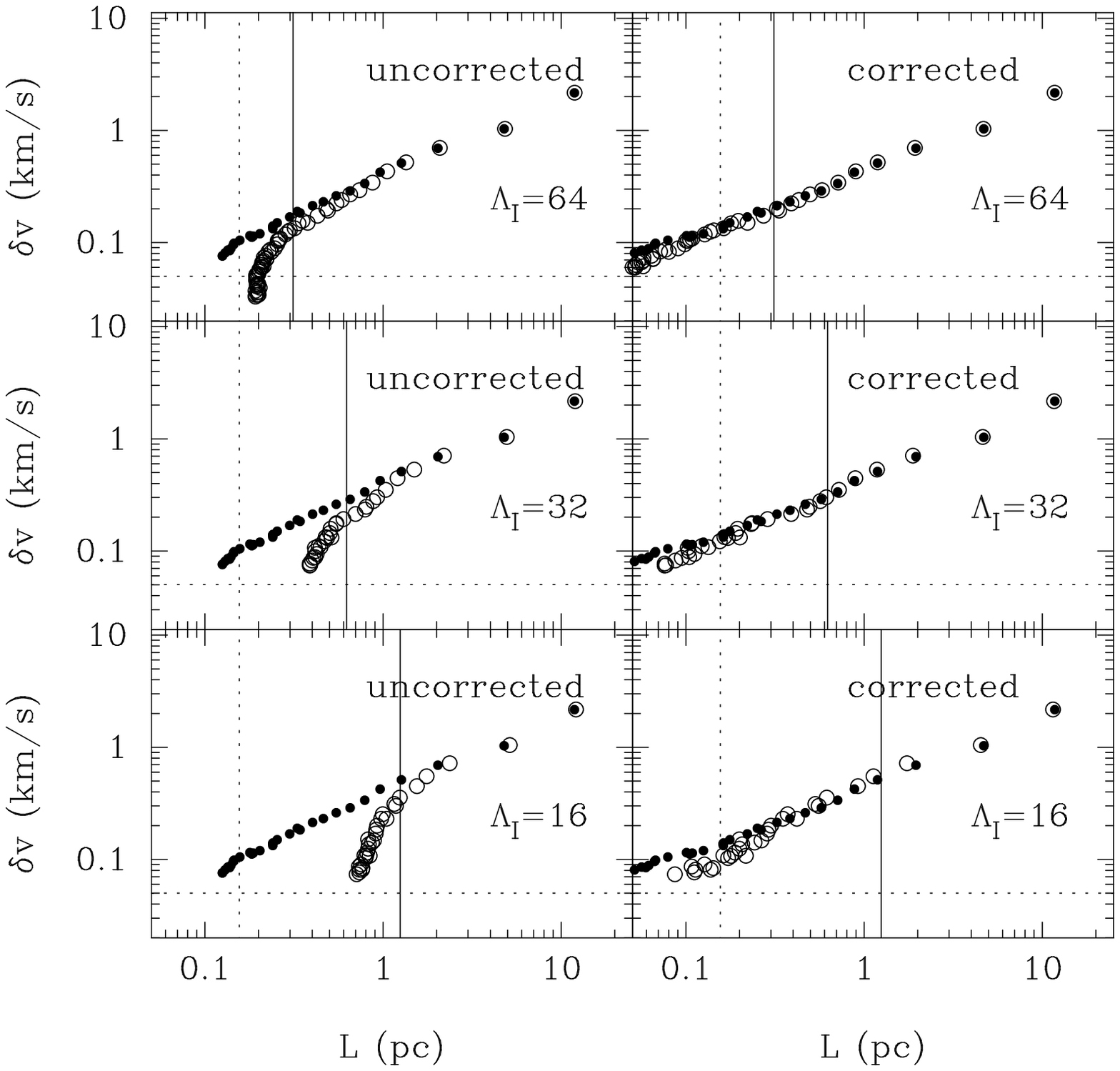}
\caption
{Spatial resolution effects on the PCA measurements. The $\lambda_{s}=128$
measurements are shown as filled circles, and the labeled
lower resolution versions as open circles. The dotted lines mark
the "full" resolution limits and the solid line the lower
resolution limits. }
\end{figure}

\begin{figure}
\plotone{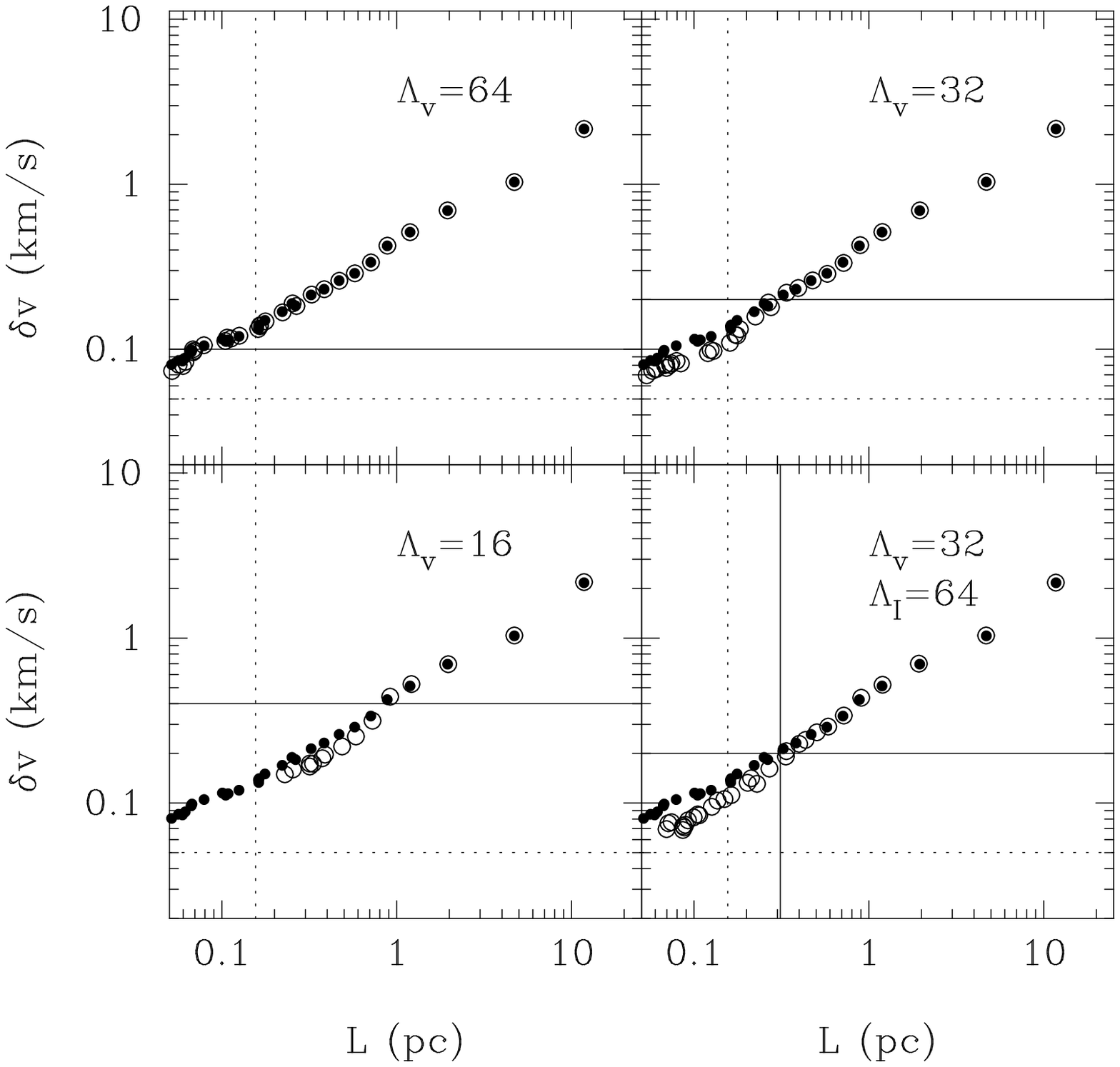}
\caption
{Spectroscopic resolution effects on the PCA
measurements. The $\lambda_{v}=128$
measurements are shown as filled circles, and the labeled
lower resolution versions as open circles. The dotted lines mark
the "full" resolution limits and the solid line the lower
resolution limits. }
\end{figure}

\clearpage
\begin{table}[htb]
\begin{center}
\caption{Uniform Density Field}
\vspace{7mm}
 \begin{tabular}
{cccccc}
\hline
$\beta$ & $\gamma={{\beta-1}\over{2}}$ & $\gamma_E$ & $<\gamma>$ &$\alpha$ \\
 \hline
1.0 & 0.0 & 0.00$\pm$0.03 & -0.02$\pm$0.04 &0.30$\pm$0.06   \\
1.5 & 0.25 & 0.26$\pm$0.02 & 0.23$\pm$0.04 &0.45$\pm$0.03 \\
2.0 & 0.50 & 0.50$\pm$0.01 & 0.48$\pm$0.05 &0.62$\pm$0.02  \\
2.5 & 0.75 & 0.72$\pm$0.01 & 0.73$\pm$0.04 &0.79$\pm$0.04  \\
3.0 & 1.00 & 0.85$\pm$0.02 & 0.86$\pm$0.05 &0.92$\pm$0.04  \\
4.0 & 1.50 & 0.93$\pm$0.02 & 0.94$\pm$0.01 &1.02$\pm$0.06  \\
\hline
 \end{tabular}
 \end{center}
\label{gammastat}
\end{table}

\begin{table}[htb]
\begin{center}
\caption{Lognormal Density Fields}
\vspace{7mm}
 \begin{tabular}
{cccc}
\hline
$\beta$ & & $\alpha$ & \\
 \hline
& $n=10^2$\cc & $n=10^3$\cc & $n=10^4$\cc \\
 \hline
1.0  & 0.40$\pm$0.01 &  0.47$\pm$0.01 & 0.44$\pm$0.01 \\
2.0  & 0.57$\pm$0.03 &  0.61$\pm$0.02 & 0.64$\pm$0.01 \\
3.0  & 0.92$\pm$0.03 &  0.97$\pm$0.03 & 0.92$\pm$0.03 \\ 
\hline
 \end{tabular}
\end{center}
\end{table}

\begin{table}[htb]
\begin{center}
\caption{Lognormal Density Field with Noise}
\vspace{7mm}
 \begin{tabular}
{cccccc}
\hline
$\beta$ & & $\alpha$ & & & \\
 \hline
& & $\zeta=\infty$ & $\zeta=4.0$ & $\zeta=2.0$ & $\zeta=1.0$ \\
 \hline
1.0 & & 0.44$\pm$0.01 &  0.45$\pm$0.01 & 0.41$\pm$0.03 &  0.48$\pm$0.01\\
2.0 & & 0.64$\pm$0.02 &  0.64$\pm$0.03 & 0.64$\pm$0.02 &  0.57$\pm$0.02\\
3.0 & & 0.92$\pm$0.03 &  0.88$\pm$0.02 & 0.81$\pm$0.01 &  0.63$\pm$0.04 \\
\hline
 \end{tabular}
\label{tabzeta}
\end{center}
\end{table}

\end{document}